\documentstyle[osa,manuscript]{revtex}
\begin{document}

\title{Electronic structure of superconducting $MgB_2$ and
related binary and ternary borides
}

\author{N.I. Medvedeva, A.L. Ivanovskii,}

\address{Institute of Solid State Chemistry, Ural Branch of the Russian
Academy of Sciences, 620219 Ekaterinburg, Russia}

\author{J.E. Medvedeva  and A.J.Freeman}

\address{Department of Physics and Astronomy, Northwestern University,
Evanston, Illinois 60208-3112}

\maketitle
\begin{abstract}

First principles FLMTO-GGA electronic structure calculations
of the new medium-$T_C$ superconductor (MTSC)
$MgB_2$ and related diborides indicate that superconductivity
in these compounds is related to the
the existence of $p_{x,y}$-band holes at the $\Gamma$ point.
Based on these calculations, we explain
the absence of medium-$T_C$ superconductivity for $BeB_2$, $AlB_2$
$ScB_2$ and $YB_2$. The simulation of a number of $MgB_2$-based ternary
systems using a supercell approach demonstrates that {\it (i)} 
the electron doping of $MgB_2$ (i.e., $MgB_{2-y}X_y$ with $X=Be, C, N, O$) 
and the creation of isoelectronic defects
in the boron sublattice (nonstoichiometric $MgB_{y<2}$)
are not favorable for superconductivity, 
and {\it (ii)} a possible way of searching for similar MTSC 
should be via hole doping of $MgB_2$ 
(i.e., $Mg_{1-x}M_xB_2$ with $M=Be, Ca, Li, Na, Cu, Zn$)
or $CaB_2$ or via creating layered superstructures of the
$MgB_2$/$CaB_2$ type. A recent report of superconductivity in $Cu$ doped
$MgB_2$ supports this view.

\end{abstract}

\newpage

The discovery of superconductivity in magnesium diboride ($MgB_2$)
\cite{c1} has attracted a great deal of interest in this system. The
transition temperature for $MgB_2$ ($T_C \approx 39$ K)
exceeds by almost two times the record values of $T_C$ for conventional $B1$-
and $A15$-type intermetallic superconductors (SC) \cite{c2}. As distinct
from the high-temperature SC, $MgB_2$ has an exclusively simple
composition and crystal structure \cite{c3}.

The finding of a new medium-$T_C$ SC (MTSC) raises a number of questions:
1. What is the nature of the superconductivity in $MgB_2$? 2. Is $MgB_2$ a
unique compound of this kind or is it the first representative of a new
class of MTSCs? 3. What physical and chemical properties should related
MTSC's possess?

It is worth noting that among boron-containing phases $MgB_2$ is not a
unique superconductor. A systematic search for superconductivity in
a wide range of $d$-metal borides (and other metal-like compounds of a
transition metals (TM) with light non-metals (carbides and nitrides
\cite{c4} )) has been carried out earlier \cite{c5,c6}. For instance,
it was shown that the superconducting transition temperature for TM
diborides ($MB_2, M = Ti, Zr, Hf, V, Ta, Cr, Mo$) is below $\sim 0.4$ K.
Only $NbB_2$ was found to be a superconductor with a $T_C$ of about 0.6 K.
A large class of superconducting (with $T_C$ of several K) ternary
($LnRuB_2$, $LnRh_4B_4$) and pseudo-ternary $(Ln_{1-x}Ln'_y)Rh_4B_4$
borides is known \cite{c7}. In 1994, Cava et al \cite{c8}  discovered
superconductivity ($T_C \approx 16-23$ K) in a new class of
intermetallic borocarbides (IBC), namely layered compounds of the
$LnM_2B_2C$ type. Numerous theoretical and experimental studies (see
review \cite{c9} ) made it possible to place IBC into the class of
conventional BCS superconductors. It is significant that bands near the
Fermi level of IBC participating in electron-phonon
interactions (with a high-frequency $Ba_{1g}$ mode) are determined by
interatomic $\sigma$-bonds in the $TMB_4$ tetrahedra. For all the
above mentioned borides, the most important role in their
superconductivity is played by $d$-electron atoms.

The electronic properties of $MgB_2$ are of quite a different kind.
Earlier band structure calculations of $MgB_2$ (by semiempirical LCAO
\cite{c10} and first principles full potential linear muffin-tin orbital
\cite{c11}, FLMTO, methods) showed that the
upper filled energy bands of $MgB_2$ are formed mainly due to strong $B-B$
interactions (in honeycomb layers of boron atoms). Analogous conclusions
were made recently by other authors \cite{c12,c13,c14,c15,c16}, who
pointed out that the coexistence of 2D in-plane and 3D interlayer bands
is a peculiar feature of $MgB_2$. Based on an estimate of phonon
frequences and band structure calculations, Kortus et al
\cite{c12} explain the superconductivity in $MgB_2$ as a result of strong
electron-phonon coupling and An and Pickett \cite{c16}  attributed it to
the behavior of $p_{x,y}$-band holes in negatively charged boron planes.
All authors \cite{c12,c13,c14,c15}  emphasize the most significant role
of metallic B states in the appearance of superconductivity.
According to the McMillan formula for $T_C$ \cite{c12}, the most probable
$MgB_2$-based MTSC should have a high density of states at the Fermi
level, $N(E_F)$, high averaged electron-ion matrix elements, 
as well as high phonon frequences, which increase
for light elements and depend on $B-B$ and $M-B$ bonding.

To examine the possibility of superconductivity in related diborides, we
studied the band structure of $MgB_2$ in comparison with that of
$CaB_2$, $BeB_2$, $AlB_2$, $ScB_2$ and $YB_2$ which are typical
representatives of different groups of $AlB_2$-like diborides formed by
$s$-, $p$-, and $d$-metals, respectively. This makes it possible to
analyze the following effects in: (1) $MgB_2$ versus $CaB_2$ and $BeB_2$,
which
 are isoelectronic and isostructural. The basic changes will be
due to structural factors (changes in lattice parameter, a, and
interatomic distance, $c/a$). (2) $MgB_2$ vs. $AlB_2$, for which the main
differences are expected as a result of changes in the filling of
bands with addition of an electron.
(3) $MgB_2$ vs.
$ScB_2$ and $YB_2$, where alongside changes in the filling of bands,
the band structure and interatomic
bonding will depend on the  M-sublattice ($s$- or
$d$-metal). (4) $AlB_2$ and $ScB_2$, $YB_2$, which are also
isoelectronic and isostructural, and changes will be due to the nature
of the $M$-sublattice ($p$- or $d$-metal).

In addition, using the supercell
approach, we have carried out a theoretical search for possible
superconductors among some $MgB_2$-based ternary systems. For this purpose, we
modelled the effect produced on the band structure of $MgB_2$ by (1)
boron sublattice doping (with $Be$, $C$, $N$, $O$ impurities), (2)
magnesium sublattice doping (with $Be$, $Ca$, $Li$, $Na$, $Cu$, $Zn$ impurities),
and (3) the presence of lattice vacancies in $Mg$- and $B$-sublattices.

The above diborides have a hexagonal crystal structure ($AlB_2$-type,
space group $P6/mmm$, $Z=1$) with atomic positions $M$ (a) $000$, $B$
(d) $1/3, 2/3,1/2$ \cite{c3}. The lattice is composed of layers of
trigonal prisms of $M$-atoms in the center of boron atoms, which form
planar graphite-like networks. Their band structures
were calculated by the FLMTO
 method within the generalized
gradient approximation for the exchange
-correlation potential
(FLMTO-GGA) \cite{c16}. Ternary and
nonstoichiometric diborides were simulated with 12-atom supercells
($2 \times 2 \times 1$) and experimental lattice parameters for  $MgB_2$,
$BeB_2$, $AlB_2$, $ScB_2$ and $YB_2$ taken from Ref. \cite{c3}. For the
hypothetical phase of $CaB_2$, the lattice constants ($a = 3.205$ \AA, $c/a
= 1.24$) were determined by total energy minimization.

$MgB_2$. The band structure and DOS are shown in
Figs. \ref{fig1} and \ref{fig2}.
The high energy part of the valence band
(VB) of $MgB_2$, made up predominantly of $B$ $2p$-states, form two
distinct sets of bands of the $\sigma$ ($2p_{x,y}$)- and $\pi$
($p_z$)-types, whose $k$ dependence differs considerably.
For $B$ $2p_{x,y}$ the most pronounced
dispersion is observed along $\Gamma-K$. These
bands are of the quasi-two dimensional (2D) type, form a flat zone
in the $k_z$ direction ($\Gamma-A$) and reflect the distribution of
$pp_\sigma$-states in the $B$ layers. These states make a considerable
contribution to $N(E_F)$, forming metallic
properties of the diboride with $E_F$ located in the region of bonding
states, the conductivity is due to hole carriers.

The $B$ $2p_z$-like bands are responsible for weaker $pp_\pi$-interactions.
These 3D-like bands have maximum dispersion along
($\Gamma-A$); $Mg$ $s,p$- and $B$ $s$-states are admixed with $B$ $2p$ bands
near the bottom of the VB and in the conduction band. Thus, the
peculiarities of the electronic properties of $MgB_2$ are associated
with metallic $2p$-states of $B$ atoms located in planar nets, which
determine the DOS in the vicinity of $E_F$.

The valence charge density (VCD) maps
(Fig. \ref{fig3}) demonstrate that
the interatomic bonds in the boron graphite-like layers
are highly covalent, the  Mg-Mg bonds have metallic character and
then there are the weak interlayer covalent $Mg-B$ interactions.
Estimates of interatomic bonds strengths (with FLMTO calculations
using empty $Mg$ and $B$ sublattices \cite{c11}) also
shows the largest contribution to the cohesion energy (Ecoh)
from $B-B$ interactions ($B-B$ (68\%), $B-Mg$ (23\%) and $Mg-Mg$ (9\%)).
In view of these results (see also \cite{c12,c13,c14,c15}), let us
discuss the possibility of superconducting properties in related
diborides.

{\it $MgB_2$ versus $CaB_2$.} The electronic bands of these isoelectronic
(electronic concentration per atom, $n_e=2.67$) and isostructural
compounds  turned out to be similar on the whole, see Fig. \ref{fig1}.
The differences are due to the increased lattice parameter, $a$,
and interlayer ($M-B$) distances determined by M ionic radii
($r_c = 0.74 (Mg)$ and 1.04 \AA\ ($Ca$)). The main change is a downward
shift of the $p_z$-bands as compared to $p_{x,y}$-bands, so that the
crossing point is located at $E_F$ (Fig. \ref{fig1}) and
$N(E_F)$ increases to the highest value of any material given in
Table \ref{tab1}. This may well indicate that a possibly higher 
T$_c$ may be obtained if $CaB_2$ were stabilized. To this end,
we also studied the lattice stability of the
hypothetical $CaB_2$ by calculating the formation energy as a difference
in the total energy with reference to the constituent elements in their
stable modifications, viz. hcp calcium and rhombohedral boron
($\alpha-B_{12}$). It was found that the formation energy has a
small, but negative value ($E_{form}=-0.12$ eV/f.u.), which may be
indicative of the possibility of real synthesis of $CaB_2$. For
comparison, $E_{form}$ for $MgB_2$ was found to be -1.21 eV/f.u..

Let us outline ways to stabilize $CaB_2$. It is known for
stable diborides that $c/a$ does not exceed $\approx 1.165$
\cite{c3}. For the equilibrium state of $CaB_2$, we obtained 
$c/a = 1.24$. The most obvious way of decreasing the interlayer distance
consists in the partial replacement of $Ca$ by atoms with smaller radii. 
To obtain superconductivity in such doped ternary systems,
these metals should be either isoelectronic with $Ca$ (for example,
$Mg$, $Be$) or hole dopants ($Li$, $Na$, etc.) $-$ as discussed below. 
They may also be prepared as layered superstructures, for example, 
$.../Ca/B_2/Mg/B_2/...$.

{\it $MgB_2$ versus $BeB_2$.} Their band structures
turned out to be similar, see Fig. \ref{fig1}. The differences
are due to a downward shift of $p_{x,y}$-bands, so that they are absent
above $E_F$ at $\Gamma$. This brings about changes
in the Fermi surface topology: cylinders along $\Gamma-A$
are transformed into cones. As a result, $BeB_2$ is not a
MTSC $-$ as supported by the
absence of superconductivity in recent experiments \cite{c16}.

{\it $MgB_2$ versus $AlB_2$ }.
For $AlB_2$ ($n_e=3.00$), the $p_{x,y}$-bands
are completely filled and $AlB_2$ is not a superconductor. A recent first
report \cite{c18}  on the electron-doped
$Mg_{1-y}Al_yB_2$, demonstrates that $T_C$ decreases
smoothly and vanishes at $y = 0.25$. According to our FLMTO estimates
\cite{c11}, the interatomic bonding exhibits some changes: the
contribution of $Al-Al$ bonds to the cohesive energy increases by almost
two times and the $M-B$ bonds become covalent (see $AlB_2$ VCD
maps in Fig. 4 of Ref. \cite{c11}).

{\it $MgB_2$ versus $ScB_2$ and $YB_2$.}  Quite a number of studies of the
electronic properties of these diborides are known to date
\cite{c19,c20,c21,c22,c23,c24}. We previously \cite{c19,c20}  performed
FLMTO calculations of all hexagonal diborides of $3d$- ($Sc,
Ti...Fe$), $4d-$ ($Y, Zr...Ru$), and $5d-$ ($La, Hf...Os$) metals and
analyzed variations in their chemical stability and some other
properties (e.g., melting temperatures, enthalpies of formation). We found
that the evolution of their band structures can be
described within a rigid-band model (RBM). For $M = Ti$, $Zr$, $Hf$
($n_e=3.0$), $E_F$ falls near the DOS minimum (pseudogap)
between the fully occupied bonding bands and unoccupied antibonding
bands. In the RBM, $MgB_2$ and $Sc$, $Y$, $La$ diborides
all have partially unoccupied bonding bands.
Comparison of their band structures shows that for $ScB_2$,
the  2D-$2p_{x,y}$
bands are almost filled and the hole concentration is very small (near
$A$ in Fig. \ref{fig1}). The $Sc$ $d$-band along $\Gamma-M$
is below $E_F$ and the large contribution to DOS at $E_F$ 
($N(E_F)$=1.06 states/eV-f.u.) is due to $Sc$ $d$-states. 
The covalent $M-B$ bonding 
increases considerably and the $B$ $2p_z$-like bands are shifted 
downwards at $K$. For both $ScB_2$ and $YB_2$ there
is a small hole concentration of $B$ $2p_{x,y}$ states at $A$.
Thus, one can expect for these diborides only low temperature 
superconductivity (for $ScB_2$ $T_C$ $\sim 1.5$ K \cite{c25}).

{\it Modeling of new $MgB_2$-based ternary borides.} In searching
for possible superconductors among the $MgB_2$-based ternary
systems, we simulated the effect produced on the band structure of
$MgB_2$ by {\it (i)} doping the B sublattice with  $C$, $N$, $O$
impurities, {\it (ii)} doping the Mg sublattice with $Be$, $Ca$, $Li$,
$Na$, $Cu$, $Zn$ impurities, and {\it (iii)} adding vacancies in the $Mg-$ and
$B-$sublattices $-$ all of which can be
divided into three groups: electron-, hole-dopants and isoelectronic
defects.

The above mentioned RBM is widely used to analyze such
substitutions. Based on the DOS of $MgB_2$ (Fig. \ref{fig2}), this model
determines the following effects: hole doping should lead to partial
"depopulation" of bonding bands, a shift of $E_F$ deep into
the VB, and an increase in $N(E_F)$. A reverse effect, namely
a Fermi level shift
to the region of the DOS minimum (pseudogap between bonding and
antibonding states), can be expected for electron doping. The role of
isoelectronic dopants remains unclear within this model. One can only
suppose that in this case the main effects will be associated with
lattice deformations, i.e., with the changes in $a$ and $c/a$.
By contrast, our FLMTO
calculations for hypothetical ternary and nonstoichiometric borides
demonstrate a more complicated picture (see Figs. \ref{fig2} and \ref{fig4}
and Table \ref{tab1}) and lead us to the following conclusions:

{\it Boron sublattice doping.} As $n_e$ decreases ($MgB_2  \rightarrow
MgB_{1.75}$), $E_F$ is shifted to higher binding energies, but
the value of $N(E_F)$ becomes smaller due to a partial breakdown of $B-B$
bonds in the graphite-like B layers and changes in the energy band
dispersion near $E_F$. For $MgB_{1.75}C_{0.25}$, all bonding
states are completely filled and $E_F$ falls in the pseudogap,
hence a decrease in $N(E_F)$. A further increase in $n_e$
($MgB_{1.75}(N,O)_{0.25}$) results in the occupation of $2p$-antibonding
states. Thus, we do not see any prospects for
obtaining superconductivity in $MgB_2$-based ternary compounds by
doping the $B$-sublattice. The considerable energy cost of breaking the strong
$B-B$ bonds should be emphasized. As reported in \cite{c11}, the
contribution of $B-B$ bonds to the cohesive energy of $MgB_2$ is 3.43
eV/atom. The corresponding values for $Mg-Mg$ and $Mg-B$ bonds are
0.45 and 1.16 eV/atom, respectively. This means that substitutions
in the $Mg$ sublattice will be energetically more preferable.

{\it Magnesium sublattice doping.}
As $n_e$ decreases in the series
 $MgB_2 \rightarrow Mg_{0.75}Li(Na)_{0.25}B_2
\rightarrow Mg_{0.75}B_2, Mg_{0.5}Li(Na)_{0.5}B_2$,
$E_F$ shifts to higher binding
energies, but no increase in $N(E_F)$ is observed $-$ as predicted
within the RBM, see Table \ref{tab1}. Thus, we found
that for $Li$ and $Na$ substitution, $N(E_F)$ is almost independent
of the hole dopant concentration.
This fact is in agreement with the results of An and Pickett \cite{c15}.
The dependence of the energy of
$\sigma(2p_{x,y})$-bands (forming cylinders along $\Gamma-A$)
at $\Gamma$ relative to $E_F$
($E^\sigma(\Gamma-E_F)$) is shown in Fig. \ref{fig4}. This dependence
for both the hole and electron doping of $MgB_2$ is close to linear. It
is seen that hole doping in $MgB_2$-based ternary systems
serves the peculiarities of band structure of $MgB_2$ near the $E_F$. 
Moreover, both $N(E_F)$ and
$E^\sigma(\Gamma-E_F)$ in these systems are practically independent of
the hole dopant type, and are determined only by the hole concentration (see
the results for
$Mg_{0.75}Li_{0.25}B_2$ and $Mg_{0.75}Na_{0.25}B_2$ or for
$Mg_{0.5}Li_{0.5}B_2$, $Mg_{0.5}Na_{0.5}B_2$ and $Mg_{0.75}B_2$ in Table
\ref{tab1} and Fig. \ref{fig1}).
For the isoelectronic substitution $MgB_2 \rightarrow 
Mg_{0.75}Be(Zn)_{0.25}B_2$,
the value of $N(E_F)$ decreases but $p_{x,y}$-bands
move up by 0.1 eV at $\Gamma$ comparing with  $MgB_2$, Fig. \ref{fig4}.

Very interestingly, in a recent experiment \cite{c26} $T_C$ was found 
to be increased up to 49 K with nominal 20 \% Cu doping of $MgB_2$.
In our calculation, we found that 25 \% Cu substitution
for Mg leads to an increase of $N(E_F)$ (c.f., Table \ref{tab1})
and here the top of $p_{x,y}$-bands at $\Gamma$ 
has the same location as for other doped systems with the same
hole concentration (c.f., Fig. \ref{fig4}).
This adds strong support to the validity of the systemetic results
presented here.

We believe that further searches of new MTSC should be made
via hole doping of $MgB_2$ (or $BeB_2$, $CaB_2$) or creating 
layer superstructures of the $MgB_2/CaB_2$ type.
We are continuing these investigations taking into account the crystal
lattice relaxation and calculating the substitution energies
for impurities in ternary systems.

We thank V. Antropov, J. Jorgensen and D. Novikov
for useful discussions. 
Work at Northwestern University supported by the U.S. Department of Energy
(Grant No. DE-F602-88ER45372).

 \begin{figure}
 \caption{Band structures of (a) $MgB_2$, (b) $BeB_2$, (c) $ScB_2$,
 (d) $CaB_2$, (e) $AlB_2$, and (f) $YB_2$. \label{fig1}}
 \end{figure}

 \begin{figure}
 \caption{Density of states  $MgB_2$, $CaB_2$, 
 $Mg_{0.75}Na_{0.25}B_2$ and $Mg_{0.75}Cu_{0.25}B_2$.
 \label{fig2}}
 \end{figure}

 \begin{figure}
 \caption{The valence charge density (VCD) maps for $MgB_2$ in in-plane
 ($B-B$, $Mg-Mg$) and  interlayer ($Mg-B$) sections.
 \label{fig3}}
 \end{figure}

 \begin{figure}
 \caption{
The energy of the $\sigma(2p_{x,y})$-band at $\Gamma$ relative to
$E_F$ ($E^\sigma (\Gamma-E_F)$) as a function of the hole- or
electron-type doping ($n_e$ -- the difference of valence electron
concentration relative to $MgB_2$).
 \label{fig4}}
 \end{figure}

 \begin{table}
 \caption{Density of states at the Fermi level ($N(E_F)$,
 in states/(eV-f.u.)) for $MgB_2$ and some related
 borides.\label{tab1}}

 \begin{tabular}{ldld}
 System            & $N(E_F)$ & System                & $N(E_F)$ \\
 \tableline
$MgB_2$	              & 0.73  & $Mg_{0.75}Li_{0.25}B_2$	  & 0.73 \\
$CaB_2$	              & 0.92  & $Mg_{0.5}Li_{0.5}B_2$     & 0.75 \\
$BeB_2$	              & 0.47  & $Mg_{0.75}Na_{0.25}B_2$	  & 0.73 \\ 
$MgB_{1.75}C_{0.25}$  & 0.56  & $Mg_{0.5}Na_{0.5}B_2$     & 0.76 \\
$MgB_{1.75}$	      & 0.61  & $Mg_{0.75}Cu_{0.25}B_2$   & 0.89 \\
$Mg_{0.75}B_2$        & 0.74  & $Mg_{0.75}Be_{0.25}B_2$	  & 0.62 \\
                      &       & $Mg_{0.75}Zn_{0.25}B_2$   & 0.51 \\
\end{tabular}
 \end{table}

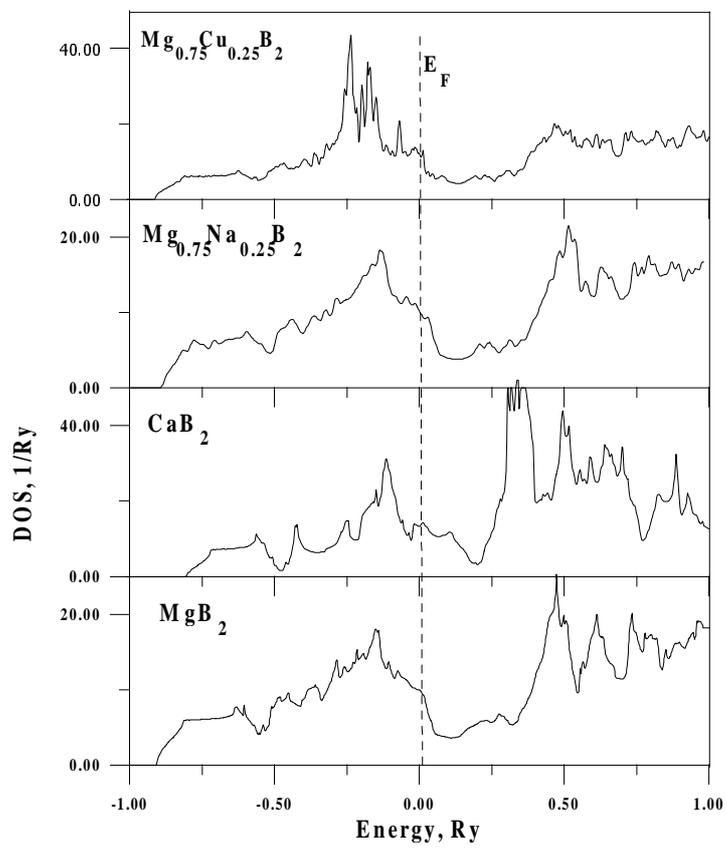
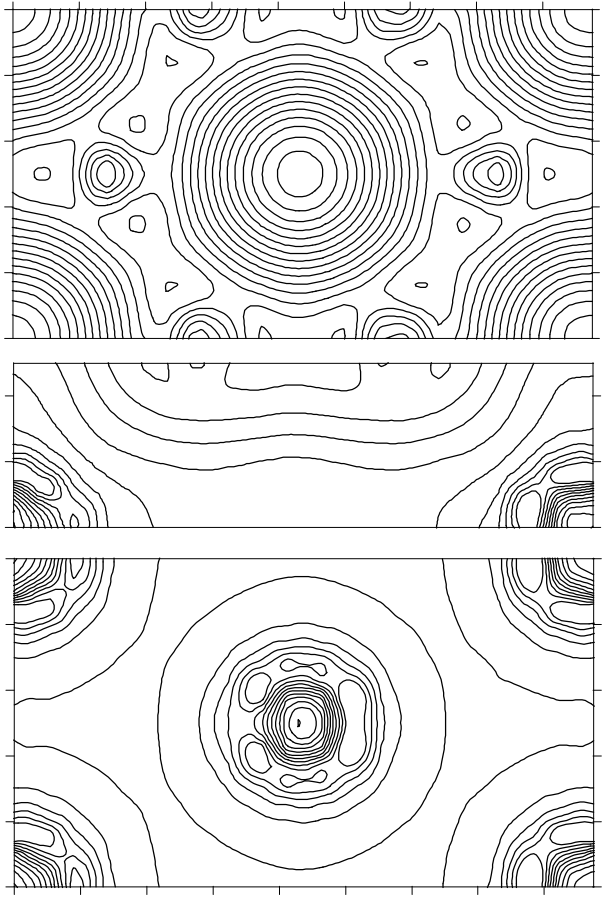
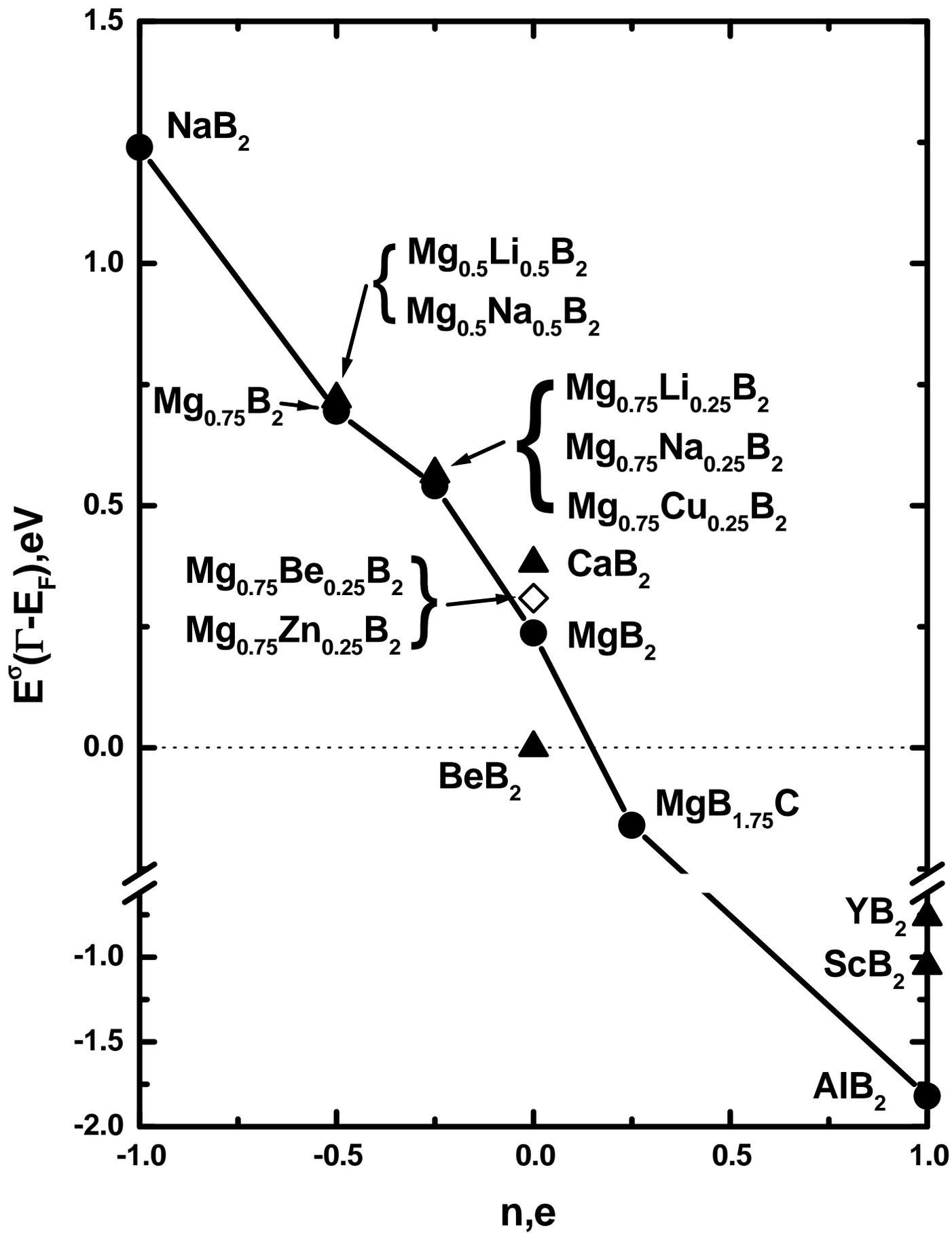
\end{document}